\documentclass{pssbmac}

\usepackage[english]{babel} 

\usepackage[utf8]{inputenc} 

\usepackage{listings}
\usepackage{xcolor}
\usepackage[colorlinks,linkcolor=blue]{hyperref}

\usepackage[T1]{fontenc}
\usepackage{float}
\usepackage{graphics}
\usepackage{graphicx}
\usepackage{epsfig}
\usepackage{indentfirst}
\usepackage{amsmath, amsfonts, amssymb, amsthm, mathtools}
\usepackage{url}
\usepackage{csquotes}

\usepackage[most]{tcolorbox}

\newtcbtheorem[auto counter,number within=section]{theo}%
  {Algorithm}{fonttitle=\bfseries\upshape, fontupper=\slshape,
     arc=0mm, colback=blue!5!white,colframe=blue!75!black}{theorem}

\usepackage[backend=biber, style=ieee, maxnames=50]{biblatex}
\DeclareTextFontCommand{\emph}{\boldmath\bfseries}
\DefineBibliographyStrings{english}{mathesis = {Master dissertation}}

\begin{document}


\title{Implementation of Kalman Filter Approach for Active Noise Control by Using MATLAB: Dynamic Noise Cancellation}

\author{
    {\large Guo Yu}\thanks{e220059@e.ntu.edu.sg} 
}
\criartitulo


\begin{abstract}
This article offers an elaborate description of a Kalman filter code employed in the active control system. Conventional active noise management methods usually employ an adaptive filter, such as the filtered reference least mean square (FxLMS) algorithm, to adjust to changes in the primary noise and acoustic environment. Nevertheless, the slow convergence characteristics of the FxLMS algorithm typically impact the effectiveness of reducing dynamic noise. Hence, this study suggests employing the Kalman filter in the active noise control (ANC) system to enhance the efficacy of noise reduction for dynamic noise. The ANC application effectively utilizes the Kalman filter with a novel dynamic ANC model.  The numerical simulation revealed that the proposed Kalman filter exhibits superior convergence performance compared to the FxLMS algorithm for handling dynamic noise. 
The code is available on \href{https://github.com/ShiDongyuan/Kalman_Filter_for_ANC.git}{GitHub} and \href{https://www.mathworks.com/matlabcentral/fileexchange/159311-kalman-filter-for-active-noise-control}{MathWorks}. 

\end{abstract}

\section{Introduction}

Active Noise Control (ANC) represents an advanced methodology for mitigating undesirable acoustic phenomena through the deployment of controlled anti-noise signals, which neutralize the primary noise via destructive interference—a nuanced application of the superposition principle within wave physics~\cite{lam2021ten,shi2023active1}. The effectiveness of ANC in attenuating low-frequency noise, coupled with its compact form factor, has facilitated its widespread adoption across various domains where noise interference poses significant challenges, including but not limited to, headphones~\cite{shen2023implementations,shen2022multi,shen2022hybrid,shen2022adaptive,shen2021alternative,shen2021wireless}, automotive industries, and architectural acoustics~\cite{lam2018active,lam2020active, lam2020active1,lam2023anti,shi2017algorithmsC,hasegawa2018window,shi2017understanding,lai2023robust,shi2016open}.

With the development of the adaptive filter theory, many adaptive algorithms have gradually evolved to be applied in the active noise control field. These derivative algorithms empower the ANC system with the ability to adapt to the variations of the noise and acoustic environments~\cite{shi2016systolic,shi2019two,wen2020convergence,shi2019practical,luo2022implementation,shi2017multiple,shi2021block,shi2019analysis,shi2020active,shi2020multichannel,shi2023multichannel,shi2023computation2,shi2024behindN, ji2023computation,shi2018novel,lai2023mov,lai2023real,shi2021comb,shi2023frequency,luo2023performance,shi2019optimal,shi2021optimal}. Moreover, because these algorithms usually aim to realize optimal control, the adaptive ANC system also has the potential to achieve the best noise reduction. During these algorithms, the filtered reference least mean square (FxLMS) algorithm plays a central role in implementing the real-time ANC system due to its high computational efficiency. Its derivative algorithms also prevail in practical applications to solve different engineering problems.  

However, these active control algorithms based on the least mean square (LMS) class still encounter an inherent issue: their sluggish convergence impacts the adaptive ANC system's ability to effectively reduce dynamic noise. The slow convergence of the ANC system also impacts consumers' perception of rapidly changing noises. Although some modified algorithms improve the response speed of the ANC system to quickly varying noise~\cite{luo2023delayless,luo2023gfanc,luo2023deep,ji2023practical,luo2022hybrid,shi2022selective,shi2023transferable,shi2021fast,shen2023momentum,shi2020feedforward}, they are still feeble to cope with some dynamic and non-stationary noises. 

The Kalman filter is known for its superior ability to track dynamic and non-stationary noise compared to traditional adaptive filter methods. Therefore, it appears to be a highly suitable choice for enhancing the convergence performance of the adaptive ANC system. However, the Kalman filter cannot be directly applied in ANC applications without an appropriate model. Hence, this study presents an innovative active noise control (ANC) dynamic model specifically designed for the Kalman filter. Moreover, the altered ANC structure is employed to implement this Kalman filter technique. The numerical simulation results confirm the efficacy of the proposed strategy. The proposed method exhibits significantly accelerated convergence speed in attenuating dynamic noise in comparison to the conventional FxLMS algorithm.

\section{Filtered-x least mean square (FxLMS) algorithm}
\label{Chapter_2_FxLMS}
\begin{figure}[htbp]
  \centering
    \includegraphics[width=11cm]{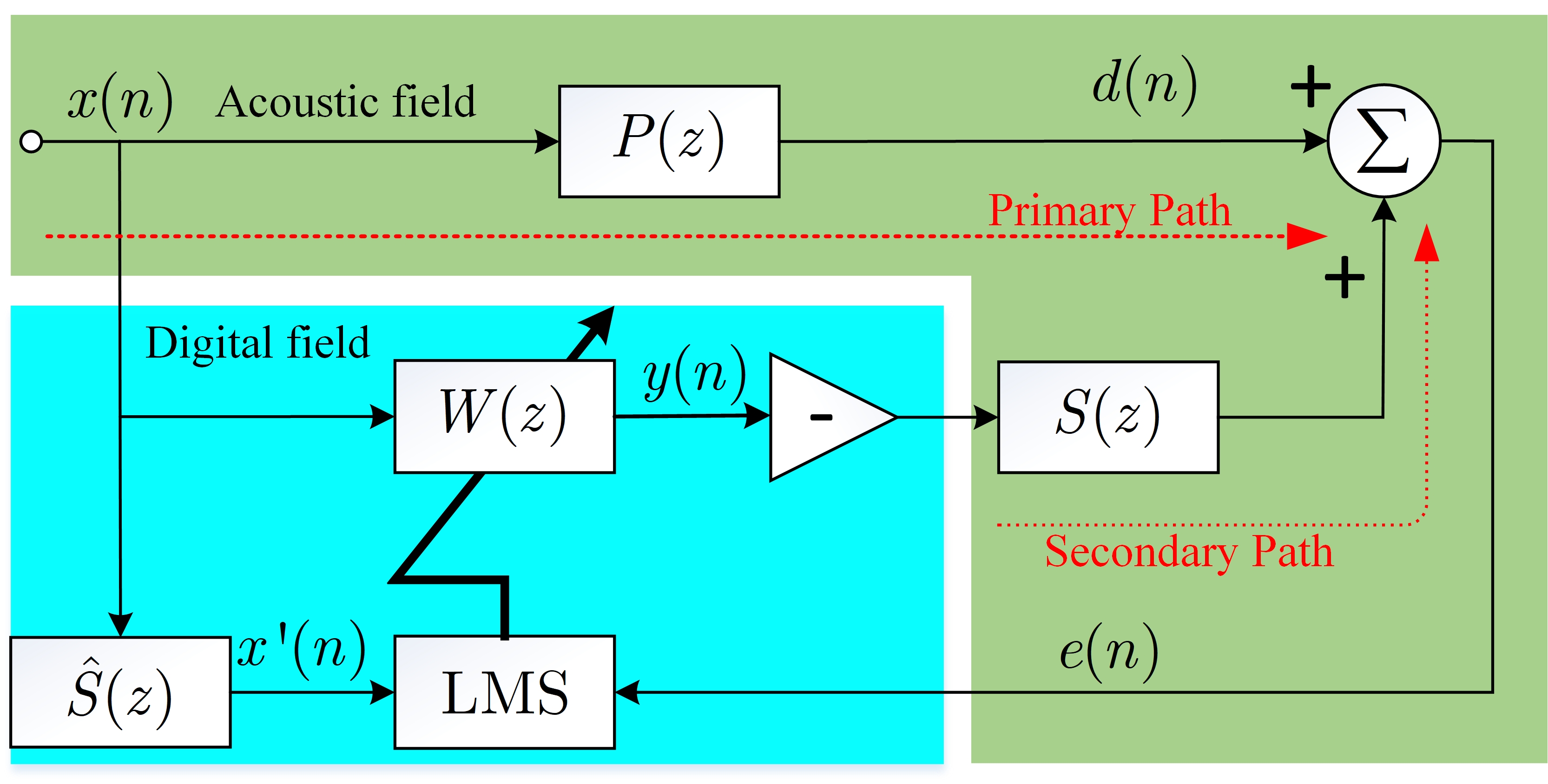}
  \caption{Block diagram of the FxLMS algorithm applied in a single-channel feedforward ANC system. (1) Green color denotes the acoustic field, and (2) Blue color denotes the digital field~\cite{shi2020algorithms}.}
  \label{Fig2_3}
\end{figure}

The filtered-x least mean square (FxLMS) is among the most practical adaptive algorithms proposed to compensate for the influence of the secondary path in an ANC system. This section introduces the FxLMS algorithm used in feedforward single- and multichannel ANC applications.
The block diagram of FxLMS algorithm is illustrated in Fig.~\ref{Fig2_3}. In the figure, $P(z)$ denotes the transfer function of the primary path from the reference microphone to the error microphone; $W(z)$ represents the control filter; $S(z)$ stands for the transfer function of the secondary path from the secondary source to the error microphone, and its estimation is $\hat{S}(z)$.     

The vector $\mathbf{x}(n)$ represents the reference signal and is defined as $\left[x(n),x(n-1),\cdots,x(n-N+1)\right]$. Let $\mathrm{T}$ denote the transpose operation and $N$ represent the length of the control filter. Therefore, the resultant output of the control filter is
\begin{equation}\label{eq2_15}
    y(n)=\mathbf{w}^\mathrm{T}(n)\mathbf{x}(n)
\end{equation}
where, $\mathbf{w}(n)$ represents the coefficient vector of the control filter. The residual error signal can be expressed as 
\begin{equation}\label{eq2_16}
    e(n) = d(n)-s(n)\ast\left[\mathbf{w}^\mathrm{T}(n)\mathbf{x}(n)\right]
\end{equation}
where, $d(n)$ and $s(n)$ denote the primary disturbance and the impulse response of secondary path, respectively; $\ast$ is the linear convolution. To minimize the instantaneous squared error, $J(n)=e^2(n)$, the most widely used method to achieve this objective is the gradient descent method, which updates the coefficient vector in the negative gradient direction with step size $\mu$:
\begin{equation}\label{eq2_17}
    \mathbf{w}(n+1)=\mathbf{w}(n)-\frac{\mu}{2}\nabla J(n)
\end{equation}
where $\nabla J(n)$ denotes the instantaneous estimate of the MSE gradient at time $n$, and can be expressed as 
\begin{equation}\label{eq2_18}
    \nabla J(n)=\nabla e^2(n)=-2\mathbf{x}'(n)e(n).
\end{equation}
The filtered reference signal vector is given by 
\begin{equation}\label{eq2_19}
    \mathbf{x}'(n)=\hat{s}(n)\ast\mathbf{x}(n)
\end{equation}
where we replace the impulse response $s(n)$ with its estimate $\hat{s}(n)$. Hence, substituting (\ref{eq2_18}) and (\ref{eq2_19}) into (\ref{eq2_17}) yields the update equation of the FxLMS algorithm
\begin{equation}\label{eq2_20}
    \mathbf{w}(n+1) = \mathbf{w}(n) + \mu e(n)\mathbf{x}'(n).
\end{equation}
If the group delay of the secondary path is assumed to be $D_\mathrm{s}$, the bound of the step size $\mu$ in the FxLMS algorithm is given by
\begin{equation}\label{eq2_bound}
    0<\mu<\frac{1}{\lambda_\mathrm{max}D_\mathrm{s}}
\end{equation}
where $\lambda_\mathrm{max}$ represents the maximum eigenvalue of the auto-correlation matrix $R_\mathrm{\mathbf{x}'}=\mathbb{E}[\mathbf{x'}^\mathrm{T}(n)\mathbf{x'}(n)]$ of the filtered reference signal. 

\section{Kalman Filter Approach for Active Noise Control}
\begin{figure}[H]
    \centering
    \includegraphics[width=10cm]{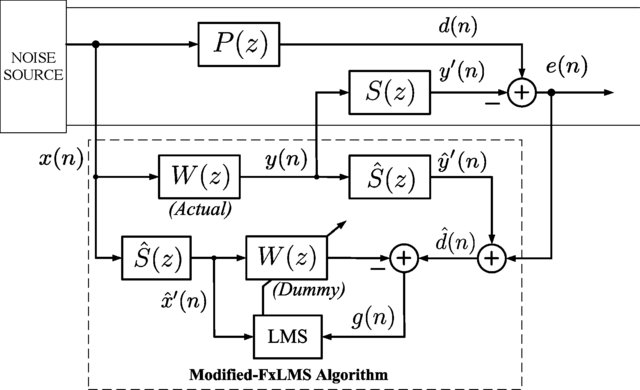}
    \caption{Block diagram of the modified ANC structure based on the FxLMS algorithm~\cite{akhtar2006new}}
    \label{fig_Modified_FxLMS}
\end{figure}

Another renowned approach for adaptive active noise reduction is the modified FxLMS algorithm, depicted in Figure~\ref{fig_Modified_FxLMS}. The internal model technique utilizes the secondary path estimate to recover the disturbance from the error signal, enabling the use of the traditional least mean square (LMS) algorithm for noise control. This altered configuration can also be employed to implement the Kalman filter methodology, wherein the LMS model is substituted with the Kalman filter to complete the updating of the control filter.

In order to employ the Kalman filter methodology, the initial step entails establishing the state function. The subsequent paragraphs will elucidate the state function definition of the ANC system. As we know, when the adaptive algorithm converges, the ANC system should achieve the optimal control filter, which is a constant solution:
\begin{equation}\label{eq_k1}
    \mathbf{w}(n+1)=\mathbf{w}(n)=\mathbf{w}_{\text{o}}, 
\end{equation}
 where $\mathbf{w}_{\text{o}}$ denotes the optimal control filter. Meanwhile, the attenuated noise can be expressed as 
 \begin{equation}\label{eq_k2}
     e_\mathrm{o}(n)=d(n)-\sum^{L-1}_{i=0}\hat{s}_l\mathbf{x}^\mathrm{T}(n-i)\mathbf{w}_\mathrm{o}(n),
 \end{equation}
where $d(n)$ and $\mathbf{x}^\prime(n)$ represent the disturbance signal and the reference vector, respectively, and $\hat{s}$ stands for the $i$-th coefficient of the secondary path estimate. Since the optimal control filter is a constant vector, \eqref{eq_k2} can be rewritten to
\begin{equation}\label{eq_k3}
    d(n)={\mathbf{x}^\prime}^\mathrm{T}(n)\mathbf{w}_\mathrm{o}(n)+e_\mathrm{o}(n),
\end{equation}
and the filtered reference signal is given by 
\begin{equation}
    \mathbf{x}^\prime(n)=\sum^{L-1}_{i=0}\hat{s}_i\mathbf{x}(n-i).
\end{equation}
\begin{figure}
    \centering
    \includegraphics[width=12cm]{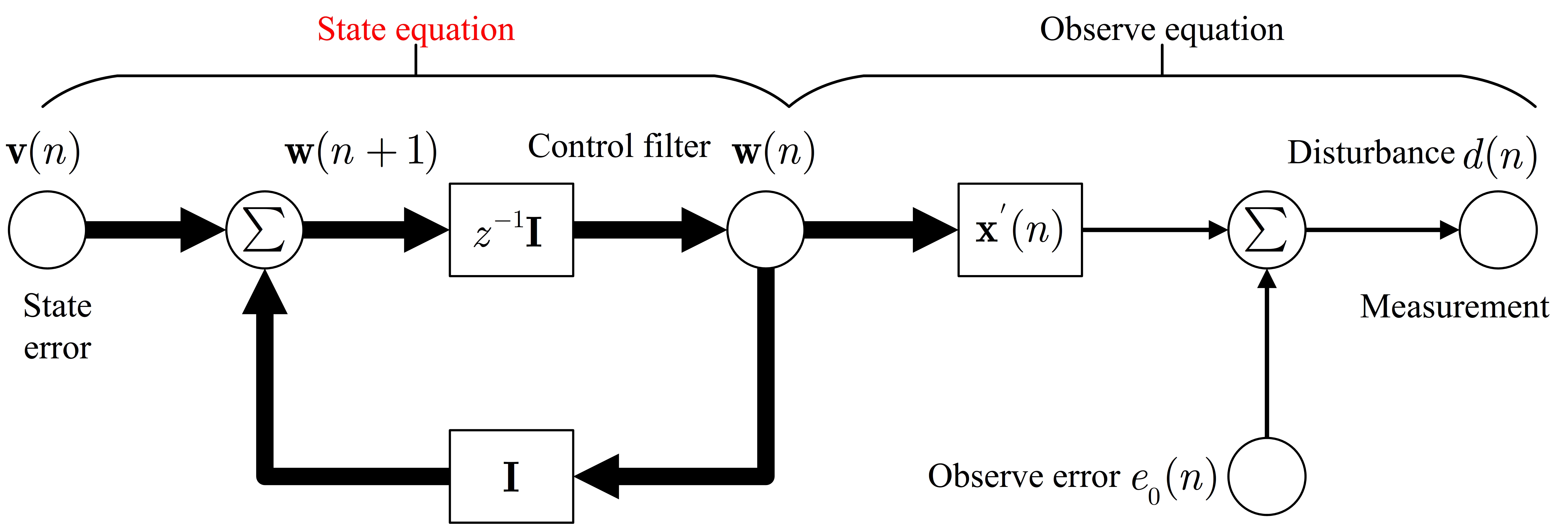}
    \caption{Signal-flow graph representation of a linear, discrete-time dynamic ANC model}
    \label{fig_kf}
\end{figure}
It is natural to let \eqref{eq_k1} and \eqref{eq_k3} be the state equation and the observation equation, respectively, as shown in Figure~\ref{fig_kf}. The Kalman filter recursive equations are listed as 
\begin{enumerate}
    \item The prediction for the new state: 
    \begin{equation}
        \hat{\mathbf{w}}(n)=\mathbf{w}(n-1)\in\mathbb{R}^{N\times 1}.
    \end{equation}
     where $N$ denotes the length of the control filter.
    \item The prediction of the auto-correlation matrix of the state error:
    \begin{equation}
        \mathbf{P}(n,n-1)=\mathbf{P}(n-n)\in\mathbb{R}^{N\times N}.
    \end{equation}
    Here, we assumed that the variance of the final state error equals $0$.

    \item The Kalman gain matrix is obtained from 
    \begin{equation}
        \mathbf{K}(n)=\mathbf{P}(n,n-1)\mathbf{x}^\prime(n)\left[\mathbf{x}^\prime(n)^\mathrm{T}\mathbf{P}(n,n-1)\mathbf{x}^\prime(n)+q(n)\right]^{-1},
    \end{equation}
    where $q(n)$ denotes the variance of the observe error:
    \begin{equation}
        q(n)=\mathbb{E}[e^2_\mathrm{o}(n)].
    \end{equation}
    \item The estimate of the state is given by 
    \begin{equation}
        \mathbf{w}(n)=\hat{\mathbf{w}}(n)+\mathbf{K}(n)\left[d(n)-\mathbf{x}^\prime(n)^\mathrm{T}\hat{\mathbf{w}}(n)\right].
    \end{equation}
    \item The auto-correlation matrix of the state error:
    \begin{equation}
        \mathbf{P}(n)=\left[\mathbf{I}-\mathbf{K}(n)\mathbf{x}^\prime(n)^\mathrm{T}\right]\mathbf{P}(n,n-1).
    \end{equation}
\end{enumerate}

Therefore, the whole Kalman filter algorithm can be summarized as 
\begin{theo}{Kalman filter algorithm}{}
    \begin{itemize}
        \item \textbf{The predicted novel state:}
        \begin{equation}
        \hat{\mathbf{w}}(n)=\mathbf{w}(n-1) \notag
        \end{equation}

        \item \textbf{The predicted state-error auto-correlation matrix:}
        \begin{equation}
            \mathbf{P}(n,n-1)=\mathbf{P}(n-n)\notag
        \end{equation}

        \item \textbf{Kalman gain:}
        \begin{equation}
            \mathbf{K}(n)=\mathbf{P}(n,n-1)\mathbf{x}^\prime(n)\left[\mathbf{x}^\prime(n)^\mathrm{T}\mathbf{P}(n,n-1)\mathbf{x}^\prime(n)+q(n)\right]^{-1}\notag
        \end{equation}

        \item \textbf{The estimated state:}
            \begin{equation}
        \mathbf{w}(n)=\hat{\mathbf{w}}(n)+\mathbf{K}(n)\left[d(n)-\mathbf{x}^\prime(n)^\mathrm{T}\hat{\mathbf{w}}(n)\right]\notag
    \end{equation}

    \item \textbf{The state-error auto-correlation matrix:}
    \begin{equation}
        \mathbf{P}(n)=\left[\mathbf{I}-\mathbf{K}(n)\mathbf{x}^\prime(n)^\mathrm{T}\right]\mathbf{P}(n,n-1).\notag
    \end{equation}
    \end{itemize}
\end{theo}

It should be emphasized that the algorithm assumes a variance of $0$ for the state error, while still accounting for the observed error. This implies that the control filter's transition has a greater level of confidence compared to the observation function. Certainly, the user can also modify the two variations while controlling the process, based on the particular application.   
\section{Code Explanation}
The section provides a concise introduction to the $KF.mat$ file, which implements the Kalman filter method for a single-channel active noise control (ANC) application. Furthermore, the FxLMS algorithm is conducted as a comparative analysis. The Kalman filter technique employs the modified feed-forward active noise control (ANC) structure, whereas the FxLMS algorithm uses the conventional feed-forward ANC structure.    

\subsection{Cleaning the memory and workspace}
This segment of code is utilized to clean the memory and workspace of the MATLAB software. 
\begin{lstlisting}[language=Matlab]
close all ;
clear     ;
clc       ;
\end{lstlisting}

\subsection{Loading the primary and secondary path}
\begin{figure}[H]
    \centering
    \includegraphics[width=12cm]{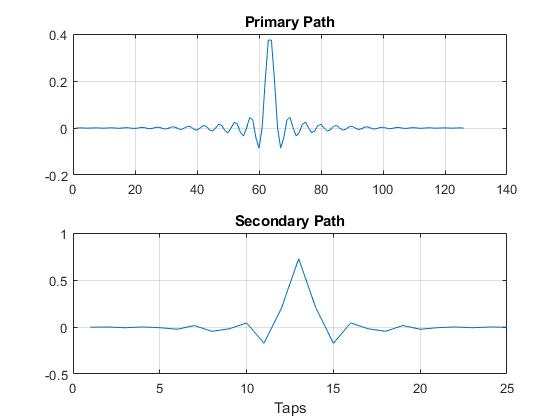}
    \caption{The impulse response of the primary path and the secondary path.}
    \label{fig_1}
\end{figure}

This part of the code loads the primary path and secondary path from the Mat files: $PriPath\_3200.mat$ and $SecPath\_200\_6000.mat$. All these paths are synthesized from the band-pass filters, whose impulse responses are illustrated in Figure~\ref{fig_1}. 

\begin{lstlisting}[language=Matlab]
load('PriPath_3200.mat');
load('SecPath_200_6000.mat') ;
figure ;
subplot(2,1,1)
plot(PriPath);
title('Primary Path');
grid on      ;
subplot(2,1,2);
plot(SecPath);
title('Secondary Path');
xlabel('Taps');
grid on      ;
\end{lstlisting}

\newpage
\subsection{Simulation system configuration}
\begin{figure}[H]
    \centering
    \includegraphics[width=12cm]{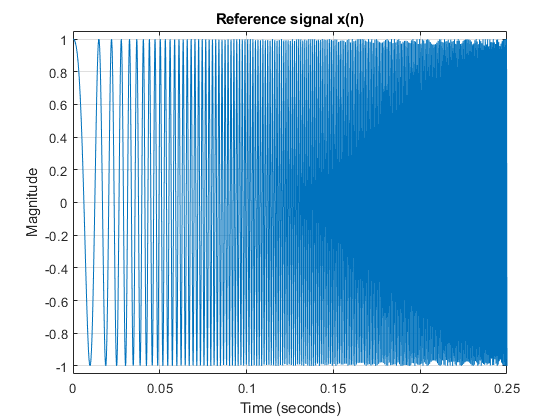}
    \caption{The waveform of the reference signal that is a chirp signal ranging from $20$ to $1600$ Hz.}
    \label{fig_2}
\end{figure}

The sampling rate of the active noise control (ANC) system is set to $16000$ Hz, and the simulation duration is $0.25$ second. To simulate the dynamic noise, the primary noise in this ANC system is a chirp signal, whose frequency gradually varies from $20$ Hz to $1600$ Hz, as shown in Figure~\ref{fig_2}.
\begin{table}[H]
\begin{tabular}{|l|l|l|l|}
\hline
\textbf{Parameter} & \textbf{Definition}       & \textbf{Parameter} & \textbf{Definition}                     \\ \hline
fs                 & Sampling rate & T                   & Simulation duration \\ \hline
y                  & Primary noise            & N                 & Simulation taps                         \\ \hline

\end{tabular}
\end{table}

\begin{lstlisting}[language=Matlab]
fs = 16000     ; % sampling rate 16 kHz.
T  = 0.25      ; % Simulation duration (seconds).
t  = 0:1/fs:T  ;% Time variable.
N  = length(t) ;
fw = 500       ;
fe = 300       ;
y = chirp(t,20,T,1600);
figure   ;
plot(t,y);
title('Reference signal x(n)');
xlabel('Time (seconds)') ;
ylabel('Magnitude')      ;
axis([-inf inf -1.05 1.05]);
grid on ;
\end{lstlisting}

\newpage
\subsection{Creating the disturbance and filtered reference}
The disturbance and filtered reference used in the ANC system are created by passing the chirp signal through the loaded primary and secondary paths.
\begin{table}[H]
\begin{tabular}{|l|l|l|l|}
\hline
\textbf{Parameter} & \textbf{Definition}       & \textbf{Parameter} & \textbf{Definition}                     \\ \hline
X                 & Reference signal vector & y                  & Primary noise \\ \hline
D                  & Disturbance vector            & PriPath                 & Primary path vector                         \\ \hline
Rf                & Filtered reference vector             & SecPath                & Secondary path vector                              \\ \hline
\end{tabular}
\end{table}
\begin{lstlisting}[language=Matlab]
%X  = 0.4*sin(2*pi*fw*t)+0.3*sin(2*pi*fe*t);
X = y;
%plot(X(end-100:end))
D  = filter(PriPath,1,X);
Rf = filter(SecPath,1,X);
%plot(D(end-100:end))
\end{lstlisting}

\subsection{Dynamic noise cancellation by the single-channel FxLMS algorithm}
\begin{figure}[H]
    \centering
    \includegraphics[width=12cm]{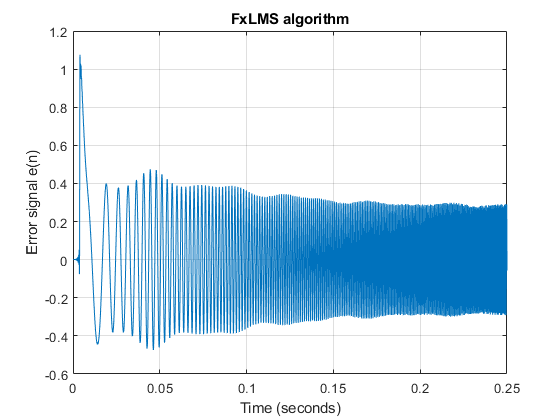}
    \caption{The error signal of the single-channel ANC system based on the FxLMS algorithm.}
    \label{fig_3}
\end{figure}

In this part, the single-channel FxLMS algorithm is used to reduce the chirp disturbance. The length of the control filter in the FxLMS algorithm has $80$ taps, and the step size is set to $0.0005$. Figure~\ref{fig_3} shows the error signal picked up by the error sensor in the ANC system. This figure shows that the FxLMS algorithm can not fully attenuate this dynamic noise during the $0.25$ second.  
\begin{table}[H]
\begin{tabular}{|l|l|l|l|}
\hline
\textbf{Parameter} & \textbf{Definition}       & \textbf{Parameter} & \textbf{Definition}                     \\ \hline
X                 & Reference signal vector & y                  & Control signal \\ \hline
D                  & Disturbance vector            & e                 & Error signal                            \\ \hline
L                & Length of the control filter              & muW                 & Step size                              \\ \hline

\end{tabular}
\end{table}
\begin{lstlisting}[language=Matlab]
L   = 80    ;
muW = 0.0005;
noiseController = dsp.FilteredXLMSFilter('Length',L,'StepSize',muW, ...
    'SecondaryPathCoefficients',SecPath);
[y,e] = noiseController(X,D);
figure;
plot(t,e) ;
title('FxLMS algorithm')   ;
ylabel('Error signal e(n)');
xlabel('Time (seconds)')   ;
grid on ;
\end{lstlisting}

\subsection{Dynamic noise cancellation by the Kalman filter approach}
\begin{figure}
    \centering
    \includegraphics[width=12cm]{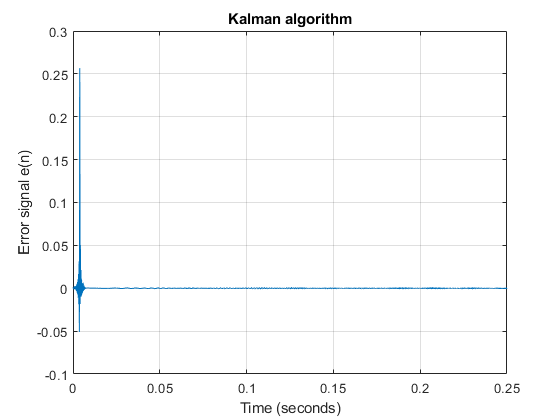}
    \caption{The error signal of the single-channel ANC system based on the Kalman filter.}
    \label{fig_4}
\end{figure}

\begin{figure}
    \centering
    \includegraphics[width=12cm]{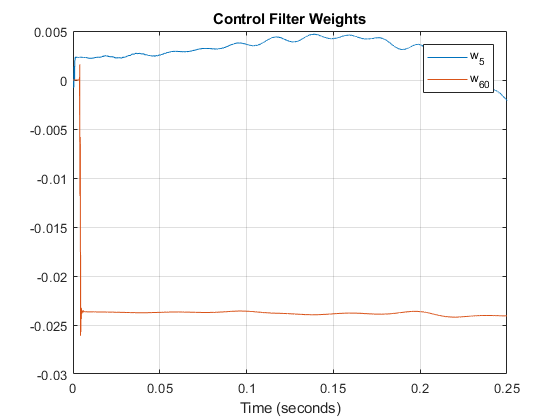}
    \caption{The time history of the coefficients $w_{5}(n)$ and $w_{60}(n)$ in the control filter.}
    \label{fig_5}
\end{figure}

\begin{figure}
    \centering
    \includegraphics[width=12cm]{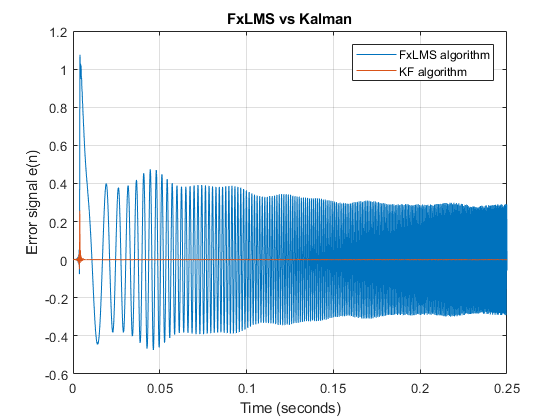}
    \caption{Comparison of the error signals in the FxLMS algorithm and the Kalman filter.}
    \label{fig_6}
\end{figure}

The Kalman filter is employed in the signal-channel ANC system to track the fluctuation of the chirp disturbance. The variance of the observed noise is initially set to $0.005$, and the auto-correlation matrix of the state error is initially set to $\mathbf{I}$, respectively. Figure~\ref{fig_4} shows the error signal of the Kalman filter algorithm. Additionally, Figure~\ref{fig_5} illustrates the variation of the coefficients $w_5(n)$ and $w_{60}(n)$ as time progresses. The outcome illustrates that the Kalman filter effectively mitigates the chirp disturbances. The Kalman filter approach has markedly superior convergence behavior compared to the FxLMS method, as illustrated in Figure~\ref{fig_6}. 

\begin{table}[H]
\begin{tabular}{|l|l|l|l|}
\hline
\textbf{Parameter} & \textbf{Definition}       & \textbf{Parameter} & \textbf{Definition}                     \\ \hline
q                  & Variance of observe error & P                  & Cross-correlation matrix of state error \\ \hline
W                  & Control filter            & ek                 & Error signal                            \\ \hline
Xd                 & Input vector              & yt                 & Anti-noise                              \\ \hline
Rf                 & Reference signal vector   &                    &                                         \\ \hline
\end{tabular}
\end{table}
\newpage
\begin{lstlisting}[language=Matlab]
q  = 0.005;
P  = eye(L);
W  = zeros(L,1);
Xd = zeros(L,1);
ek = zeros(N,1);
w5 = zeros(N,1);
w60 = zeros(N,1);

%-----------Kalman Filer---------
for ii =1:N
    Xd     =[Rf(ii);Xd(1:end-1)];
    yt     = Xd'*W ;
    ek(ii) = D(ii)-yt ;
    K      = P*Xd/(Xd'*P*Xd + q);
    W      = W +K*ek(ii)        ;
    P      =(eye(L)-K*Xd')*P    ;
    %---------------------------
    w5(ii)  = W(5);
    w60(ii) = W(60);
    %---------------------------
end
%-------------------------------
figure;
plot(t,ek);
title('Kalman algorithm')   ;
ylabel('Error signal e(n)');
xlabel('Time (seconds)')
grid on ;
figure
plot(t,w5,t,w60);
title('Control Filter Weights');
xlabel('Time (seconds)');
legend('w_5','w_{60}');
grid on ;
figure;
plot(t,e,t,ek);
title('FxLMS vs Kalman')   ;
ylabel('Error signal e(n)');
xlabel('Time (seconds)')   ;
legend('FxLMS algorithm','KF algorithm');
grid on ;
\end{lstlisting}

\section{Conclusion}
This document provides a detailed introduction to the Kalman filter code used in the active control system. Traditional active noise control typically adapts the adaptive filter, such as the filtered reference least mean square (FxLMS) algorithm, to adapt to the variations of the primary noise and acoustic environment. However, the sluggish convergence behavior of the FxLMS algorithm usually affects the noise reduction for the dynamic noise. Therefore, this work proposes using the Kalman filter in the ANC system to improve the noise reduction performance for dynamic noise. With a novel dynamic ANC model, the Kalman filter is excellently deployed in the ANC application.  The numerical simulation demonstrated that the proposed Kalman filter has a much better convergence performance than the FxLMS algorithm in dealing with dynamic noise.            
\printbibliography


\end{document}